\documentclass[aps,prd,twocolumn,groupedaddress,showpacs,amsmath,amssymb]{revtex4}
    \def\be{\begin{equation}}
    \def\ee{\end{equation}}
    \def\ba{\begin{eqnarray}}
    \def\ea{\end{eqnarray}}

    \newcommand{\ep}[1]{\epsilon_{#1}}
    \newcommand{\de}[1]{\delta_{#1}}

    \newcommand{\rd}{{\rm d}}
    
    \newcommand{\pa}[1]{\left(#1\right)}
    \newcommand{\paq}[1]{\left[#1\right]}
    \newcommand{\M}{{\rm M_{\rm P}}}
  \input epsf
\usepackage{pstcol,graphicx,latexsym}
\usepackage{graphicx}
\usepackage{graphicx,epsf}
\usepackage{epsfig}
\begin{document}
\title{Inflation and Reheating in Induced Gravity}

\author{A. Cerioni$\,^{1,2}$, F. Finelli$\,^{3,4,2}$, A. 
Tronconi$\,^{1,2}$ and G. Venturi$\,^{1,2}$}

\affiliation{$^{1}$ Dipartimento di Fisica, Universit\`a degli Studi di
Bologna, via Irnerio, 46 -- I-40126 Bologna -- Italy}
\affiliation{$^{2}$ INFN, Sezione di Bologna,
Via Irnerio 46, I-40126 Bologna, Italy}
\affiliation{
$^{3}$ INAF/IASF Bologna,
Istituto di Astrofisica Spaziale e Fisica
Cosmica di Bologna \\
via Gobetti 101, I-40129 Bologna - Italy}
\affiliation{$^{4}$ INAF/OAB, Osservatorio Astronomico di Bologna,
via Ranzani 1, I-40127 Bologna -
Italy}

\begin{abstract}
Inflation is studied 
in the context of induced gravity (IG) $\gamma \sigma^2 R$, where $R$ is the Ricci scalar, $\sigma$ a scalar field and $\gamma$ a dimensionless constant. 
We study in detail cosmological perturbations in IG and examine both a Landau-Ginzburg (LG) and a Coleman-Weinberg (CW) potential 
toy models for small field and large field (chaotic) inflation and find that
small field inflationary models in IG are constrained 
to $\gamma \lesssim 3 \times 10^{-3}$ by WMAP 5 yrs data.     
Finally we describe the regime of coherent oscillations in induced gravity
by an analytic approximation, showing how the homogeneous inflaton can decay 
in its short-scale fluctuations when it oscillates around a non-zero value $\sigma_0$.
\end{abstract}

\maketitle
Beginning with the association of the gravitational coupling with a scalar field \cite{brans} many attempts have been made to 
relate the gravitational constant to dynamics. Indeed starting from an attempt to relate it to one loop effects in some 
fundamental interaction \cite{sakharov} induced gravity (IG) theories $\gamma \sigma^{2} R$ have been developed \cite{zee,smolin,adler}. 
In such theories $\sigma$ acquires a non zero vacuum expectation value through the spontaneous breaking of scale invariance arising 
through the presence of a condensate \cite{zee} or quantum effects (radiative corrections) \cite{CW}. Further such theories can be 
generalized \cite{CV} leading to a viable dark energy model \cite{FTV}.\\ 
In IG one may also use $\sigma$ to achieve inflation. ``Old inflation'' in such a context is not satisfactory \cite{eweinberg} 
and small field and large field (chaotic) inflation appear more promising. In particular for the last case with 
$\gamma \gg 1$ the constraints on $\sigma$ are such that it may even be compatible with spontaneous symmetry breaking in the 
usual particle physics context \cite{spokoiny,SBB,FU2}.

In this Letter we analyze the slow-roll predictions for single-field inflation 
in IG and compare them with the recent WMAP 5-yrs data \cite{komatsu}. 
We also give an analytic approximate solution for the coherent oscillation regime during which reheating in IG takes place.

Let us start by considering the IG action 
\begin{equation}
S = \int d^4 x\sqrt{-g}\left[{\gamma\over 2}\sigma^2 R-\frac{g^{\mu\nu}}{2}\partial_{\mu}
\sigma\partial_{\nu}\sigma-V(\sigma) \right]
\label{original}
\end{equation}
where $\gamma$ is a dimensionless and positive definite parameter 
and we assume a spatially flat Robertson-Walker background. 
The variation of the above action leads to the following set of 
independent equations
\begin{eqnarray}
H^2 &=& \frac{1}{3\gamma \sigma^2}
\left[ \frac{\dot\sigma^2}{2}+V(\sigma) \right]-
2H\frac{\dot\sigma}{\sigma}\label{eq:Friedmann}\\
\ddot\sigma &+& 3H\dot\sigma + \frac{\dot\sigma^2}{\sigma}= - 
\frac{V_{\mathrm{eff},\sigma}}{1+6\gamma}
\label{eq:sigma}
\end{eqnarray}
where $V_{\mathrm{eff},\sigma}=\rd V/\rd \sigma-4V/\sigma$ and the dot is the derivative with respect to cosmic time.


Inflation generically occurs during the slow-rolling of $\sigma$, 
for which Eqs. (\ref{eq:Friedmann},\ref{eq:sigma}) reduce to:
\be
H^2 \simeq \frac{V}{3\gamma\sigma^2} \mbox{,}\quad
3 H \dot \sigma \simeq - \frac{V_{\mathrm{eff},\sigma}}{1+6\gamma}
\label{slowroll}
\ee
The slow-roll parameter $\epsilon_1 \equiv - \dot H/H^2$ obtains contributions not only 
from the square of the velocity of the scalar field as in EG but also from other terms:
\be
\dot H = -\frac{1}{2\gamma}\frac{\dot\sigma^2}{\sigma^2}+4H\frac{\dot\sigma}{\sigma}
+\frac{V_{\mathrm{eff},\sigma}}{\sigma(1+6\gamma)}
\ee
 
As in Einstein Gravity (EG), it is useful 
to introduce the hierarchy of Hubble flow functions \cite{STEG}: $d\ln \epsilon_n/dN = \epsilon_{n+1}$ with $n \ge 0$, $\epsilon_0 = H_i/H$ and $N$ being the number of e-folds.
In scalar-tensor gravity it is also necessary to introduce another 
hierarchy associated with $\sigma$: $d\ln \delta_n/dN = \delta_{n+1}$ with 
$n \ge 0$, $\delta_0 = \sigma/\sigma_i$. Thus in IG the cosmological perturbations depend on these two hierarchies which completely specify 
the background evolution. As usual, one can replace equations (\ref{eq:Friedmann},\ref{eq:sigma}) with an equivalent set written in terms of the two hierarchies. 
However the two hierarchies are not independent and the following relation holds:
\ba
\epsilon_1& = &\frac{\delta_1}{1+\delta_1}
\left(\frac{\delta_1}{2\gamma}+2\delta_1+\delta_2-1\right)\label{eom1}
\ea 
Inflation in IG has a richer phenomenology than in EG: inflation could occur for large $\de{1}$ and
a super-inflationary stage \cite{BFM} could take place for $|\delta_1| \ll 1$, 
($\epsilon_1 = -\dot H/H^2 < 0$) in the Jordan frame 
with $\delta_1 >0$ and $2 \delta_1 + \delta_2 + \delta_1/(2\gamma) < 1$ or viceversa.

Scalar curvature perturbations produced by quantum fluctuations of the 
inflaton during the accelerated stage are 
described by $\mathcal R(x)= - H \delta \sigma(x)/\dot \sigma$ in the uniform curvature gauge \cite{hwang}, 
where $\delta \sigma(x)$ is the scalar inflaton perturbation and is the correct field variable 
to quantize. 
The Fourier component $\delta \sigma_{k}$ of the inflaton fluctuation, 
in the IG context, has been shown to satisfy the differential equation 
\cite{hwang}:
\ba
\ddot{\delta \sigma_k}&+&\left( 3 H + \frac{\dot Z}{Z} \right)
\dot{\delta \sigma_k} + \nonumber\\
&+&\left[ \frac{k^2}{a^2} - \frac{1}{a^3 Z \sigma \delta_1}
\left( a^3 Z \left( \sigma \delta_1 \right)^\cdot \right)^\cdot
\right] \delta \sigma_k = 0\label{scperteq}
\ea
where
\be\label{Qshom}
Z = \frac{H^2 \sigma^2 (1+6\gamma)}{(\dot \sigma + H \sigma)^2} = \frac{1+6\gamma}{(1 + \delta_1)^2}.
\ee
Gravitational waves are also amplified from quantum fluctuations during inflation; in IG the Fourier modes of tensor perturbations satisfy:
\be\label{tnperteqN}
\ddot h_{s,k} + (3H + 2H\delta_1)\dot h_{s,k}+\frac{k^2}{a^2}h_{s,k}=0
\ee
where $s = + \,, \times$ denotes the two polarization states.
We define the power spectra of scalar curvature perturbations 
and tensor perturbation as
\be\label{scpowspe}
\mathcal{P}_{\mathcal{R}}(k) \equiv \frac{k^3}{2\pi^2}|\mathcal{R}_k|^2\simeq \mathcal{P}_{\mathcal{R}}(k_*) 
\left( \frac{k}{k_*} \right)^{n_s-1} 
\ee
and
\be\label{tnpowspe}
\mathcal{P}_{h}(k) \equiv \frac{2 k^3}{\pi^2}\left(|h_{+ , k}|^2
+ |h_{\times , k}|^2 \right) \,\simeq \mathcal{P}_{h}(k_*)
\left( \frac{k}{k_*} \right)^{n_t}
\ee
respectively, where $k_*$ is a suitable pivot scale.
It is important to stress that through a conformal transformation 
the action in Eq. (\ref{original}) can be rewritten as EG involving the 
rescaled scalar field and metric with a different potential: whereas the 
spectra of primordial cosmological perturbations are conformal invariant, 
we prefer to work in the original frame in Eq. (\ref{original}).

As in EG with a standard scalar field with 
exponential potential, exact inflationary solutions with
$a (t) \sim t^p$ (with $t>0$ and $p>1$) exist for
induced gravity with $V(\sigma) = \lambda_n \sigma^n/n$ and
\ba
p = \frac{1}{\epsilon_1} = 2\frac{1+ (n+2) \gamma }{(n-4)(n-2) \gamma} \,,\,\,
\, \, \sigma(t) = \frac{S}{t^\frac{2}{(n-2)}}
\label{scaling_sigma} \,,
\ea
with $4<n<4+\sqrt{2(6+1/\gamma)}$ or $4-\sqrt{2(6+1/\gamma)}<n<2$
\cite{explanation}. These are scaling solutions for which $\delta_2 = 0$ and 
\be\label{exsol}
\de{1}=-\frac{\gamma\pa{n-4}}{1+\gamma\pa{n+2}},\;\;\ep{1}=\frac{\gamma\pa{n-2}\pa{n-4}}{2+2\gamma\pa{n+2}} \,.
\ee
The poles in $n=2 \,, 4$ in the above equations
correspond to de Sitter solutions having $a(t) \propto e^{H t}$.
The above scaling solution is also found for
$a (t) \sim (-t)^p$ (with $t<0$ and $p<0$) but only for $2<n<4$:
this solution characterizes a super-inflationary stage with $\dot H > 0$
and ends up in a future singularity (the Ricci scalar grows
with the arrow of time instead of decreasing).
Let us note that inflation does not end for the above scaling solutions (just as in EG),
but cosmological perturbations can be solved exactly on such backgrounds.
The spectral index for scalar curvature perturbations is:
\be\label{exspecind}
n_{s}-1= \frac{\rd \ln \mathcal{P}_{\mathcal{R}}}{\rd \ln k} 
= \frac{2\gamma \pa{n-4}^{2}}{\gamma\pa{n-4}^{2}-2\pa{6\gamma+1}}
\ee
where $\rd n_{s}/\rd\ln k= \rd n_{t}/\rd\ln k=0$ and $n_t = n_s-1$. 
The exact tensor-to-scalar ratio is given by 
\be\label{consrel}
r=\frac{\mathcal{P}_{h}(k)}{\mathcal{P}_{\mathcal{R}}(k)}=
-\frac{8n_{t}}{1-\frac{n_{t}}{2}}
\ee
which agrees with the consistency condition of power-law inflation in EG.
On examining the behavior of (\ref{exspecind}) one has
\be
n_{s}-1<0\quad{\rm when}\quad \rho\equiv\frac{\gamma\pa{n-4}^{2}}
{2(1+6\gamma)}<1\,
\ee
and $n_{s}-1$ is close to scale invariance (zero) only for $\rho\sim0$ 
and $n_{s}-1>2$ for $\rho>1$. 
\begin{figure}
\centering
\epsfig{file=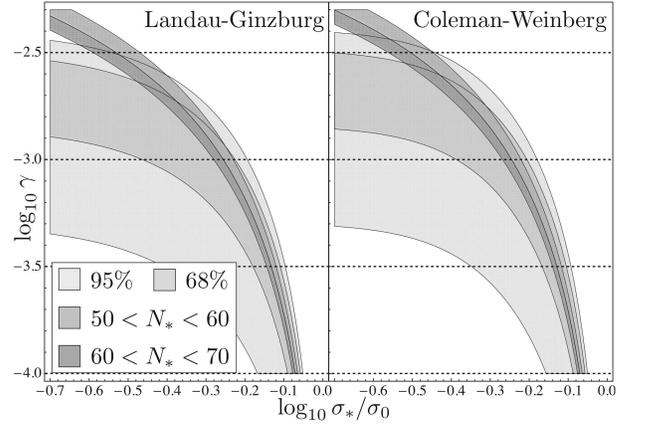, width=8 cm}
\caption{{\it The WMAP5 + BAO + SNIa constraints on $n_s$ \cite{komatsu}
in absence of gravitational waves with the condition $50< N_{*}<70$.}
\label{FIGdyn}}
\end{figure}

Let us now obtain general formulae for scalar and tensor spectra in the 
slow-roll regime. As for EG, the potential and its derivatives are given 
explicitly by the Hubble parameter and the coefficients of the 
hierarchies $(\epsilon_i, \delta_i)$. On assuming $\delta_i \ll1$, 
we can invert such relations obtaining (we only exhibit the 
first two): 
\ba
\delta_1 &\simeq& - \gamma \sigma 
\frac{V_{\mathrm{eff},\sigma}}{(1+6\gamma) \, V} \,,   
\label{delta1}
\\
\delta_2 &\simeq&  - \gamma \sigma^2 
\frac{V_{\mathrm{eff},\sigma\sigma}}{(1+6\gamma) \, V} +
\delta_1 \left( \frac{\delta_1}{\gamma} - 3 \right) \,.
\label{delta2}
\ea
From eq. (\ref{delta1}) we immediately see that the first slow-roll parameter 
in IG is roughly the square root of what we expect from EG. 
Indeed through a conformal trasformation to the Einstein frame one can show that
$\epsilon_1^{\rm EF} = \delta_1^2 (1+6\gamma)/(2 \gamma (1+\delta_1)^2)$. 
Further, from Eq. (\ref{delta1}), it is easy to see that 
$\epsilon_1 \simeq - \delta_1$ to lowest order for large $\gamma$. 
The scalar power-spectrum is 
\be\label{scampl}
\mathcal{P}_{\mathcal{R}}(k_*) \simeq
\frac{A \, H^{2}_*}{4 \pi^{2}\pa{1+6\gamma}\de{1 \, *}^{2}\sigma_*^{2}}
\simeq\frac{A \, V^3_* (1+6\gamma)}{12 \pi^2 \gamma^3 \sigma_*^6
V^{' 2}_{\mathrm{eff} \, *}}
\ee
with
\be
A = \paq{1-2\ep{1\,*}+C \pa{\de{1\,*}+\de{2\,*}+\ep{1\,*}}},
\ee
where $C=2\pa{2-\ln 2-b}$, $b$ is the Euler-Mascheroni constant and 
all the above quantities with the subscript $*$ are evaluated when $k_*$ crosses the Hubble radius. 
The scalar spectral index in Eq. (\ref{scpowspe}) is 
\ba
\!\!\!\!\!\!\!n_{s}-1 &=& -2\pa{\de{1 \, *}+\de{2 \, *}+\ep{1 \, *}} \nonumber \\
&=&\frac{2\gamma \,\sigma_{*}^{2}}{1+6\gamma}\left(\frac{V_{\mathrm{eff},\sigma\sigma *}}{V_*}-\frac{3V_{\mathrm{eff},\sigma *}^2}{2 V_*^2}-\frac{3V_{\mathrm{eff},\sigma \, *}}{\sigma_{*}V_*}\right)
\label{scspSR}
\ea
Let us note that the above result for $n_s$ 
agrees with the calculation in the Einstein frame, but does not agree 
with Ref. \cite{CY} where a term of order $\delta_{i}$ is omitted. 
We also note $V_{\mathrm{eff} \,, \sigma}$ is less than $V$ to the addition of radiative corrections of logarithmic form to the tree potential.\\
Similarly, the tensor power spectrum is given by:
\be 
\mathcal{P}_{h}(k_*) \simeq
\frac{2 (A - C \delta_{2 \, *}) H_*^{2}}{\pi^{2}\gamma\sigma_*^{2}} \simeq \frac{2 (A - C \delta_{2 \, *}) V_*}{3 \pi^{2}\gamma^2\sigma_*^{4}}
\ee
and 
\be
n_t \simeq - 2 \pa{\de{1 *}+\ep{1 *}} \simeq - 
\frac{\delta_{1 \, *}^2 (1+6\gamma)}{\gamma} \,.
\ee
The above results lead to the standard tensor-to-scalar ratio 
$r_* \simeq - 8 n_{t *}$.
We shall consider the following potentials leading to 
the spontaneous breaking of scale invariance and Newton's constant: 
\be\label{CW}
V_{\rm CW}=\frac{\mu}{8}\sigma^{4}\pa{\ln\frac{\sigma^{4}}{\sigma_{0}^{4}}-1}+\frac{\mu}{8}\sigma_{0}^{4}
\ee	
where the breaking arises through a quantum effect, Coleman-Weinberg (CW) type, and
\be\label{LG}
V_{\rm LG}=\frac{\mu}{4}\pa{\sigma^{2}-\sigma_{0}^{2}}^{2}\,.
\ee
where the breaking arises through a condensate, Landau-Ginzburg (LG) type.
\begin{figure}[h!]
\centering
\epsfig{file=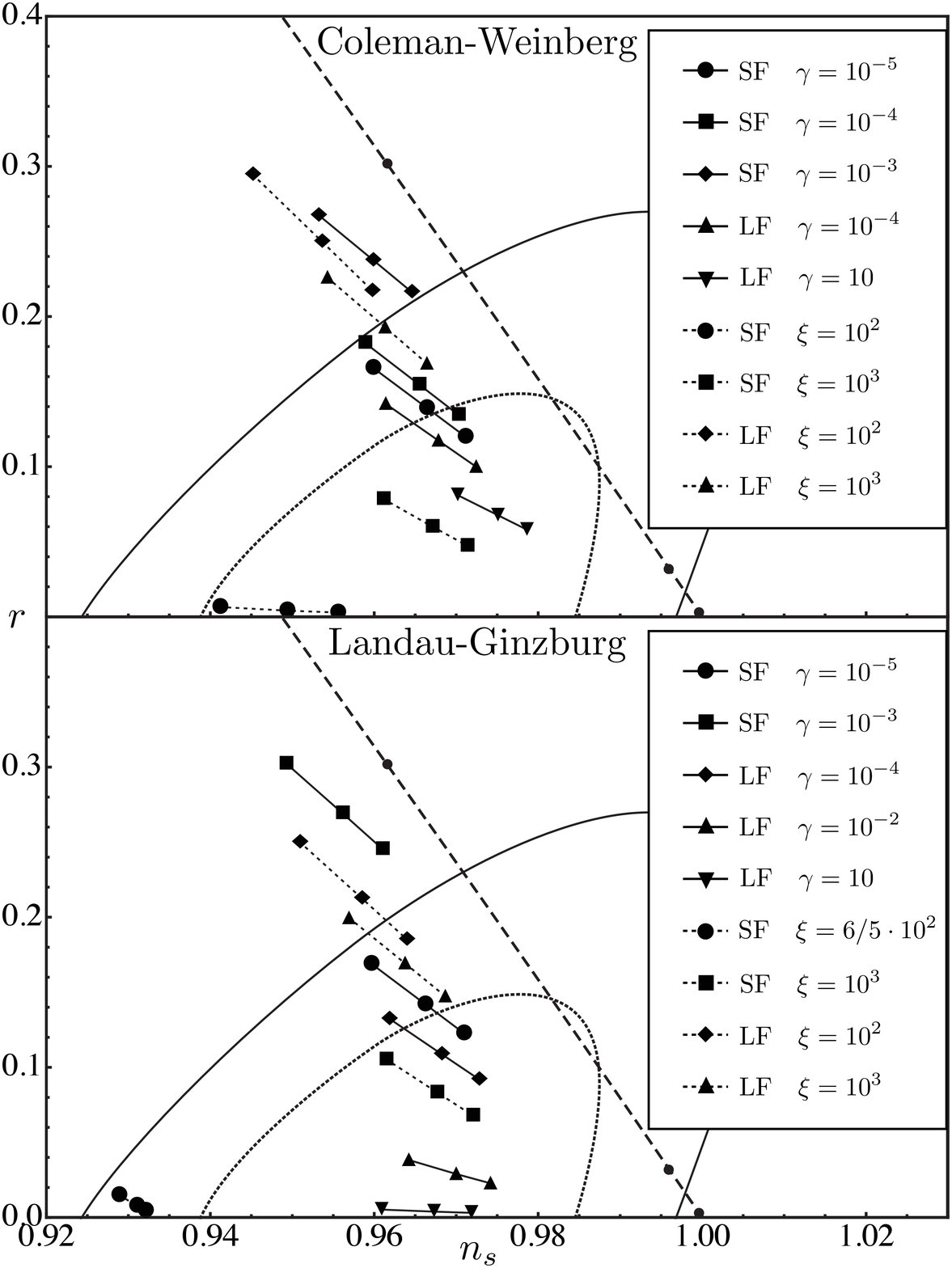, width=8.5 cm}
\caption{{\it The WMAP5+BAO+SN Ia constraints \cite{komatsu} in the $(n_s \,, r)$ plane. 
Each segment shows the uncertainty on $N_*$. The dashed line plots the consistency relation and the points on such a line are for $n=6$ and $\gamma=10^{-2},\,\,10^{-3},\,\,10^{-4}$ from the left to the right.} 
\label{rtest}}
\end{figure}
Both large and small field configurations for the potentials 
(\ref{CW},\ref{LG}) can lead to predictions in agreement with observations, 
for suitable values of $\gamma$ and the parameters of the potential, $\mu \,, \sigma_0$.
In Figure (\ref{FIGdyn}) we plotted the values of $(\sigma^{*},\gamma)$ in the small field regime and assuming that $k_{*}=a H$ for 50-70 e-folds, $N_{*}$, before 
inflation ends. Lighter regions represent the observational constraints on the spectral index $n_{s}$ coming from the WMAP 5 + BAO + SN Ia
and a $68\%$ and $95\%$ confidence level ($n_{s}=0.963\pm0.014$ and $n_{s}=0.963\pm0.028$), darker regions 
represent the two intervals $50<N_{*}<60$ and $60<N_{*}<70$ (darker area). Such a plot constraints $\gamma$ to be $\lesssim 3\cdot10^{-3}$ in order to fit observations.
We note that small field regime in IG leads to predictions very different from EG.
For both CW and LG potentials we obtain $n_{s}-1 \simeq - 16 \gamma/(1+2\gamma)$.
For small field potentials as  $V(\sigma) \propto 1 - (\sigma/\sigma_0)^n+...$ we obtain $n_{s}-1\simeq-16\gamma/(1+6\gamma)$.
For large field configurations, we obtain $n_{s}-1\simeq-2/N_{*}\,\,(n_{s}-1\simeq-1.5/N_{*})$ for LG (CW) which leaves $\gamma$ unconstrained.
The amplitude for scalar perturbations for the LG potential is 
\be
P_{\mathcal{R}}(k*)\simeq\left\{
\begin{array}{lcl}
\frac{\mu}{3\pi^2\gamma}N_*^2\mbox{,} & \gamma  \ll 1&(\mbox{large and small field})\\
\frac{\mu}{72\pi^2\gamma^2}N_*^2\mbox{,} & \gamma \gg 1&(\mbox{large field})
\end{array}
\right. \,.
\ee
WMAP5 + BAO + SN Ia require 
$P_\mathcal{R}(k^*)= (2.445\pm0.096)\times 10^{-9}$ \cite{komatsu}: 
from the above expression we observe that when $\gamma\gg1$ $\mu$ 
need not be as small as the same self-coupling in EG\cite{spokoiny,SBB,FU2}. 
For instance $\mu \sim 0.5$ is allowed for $\gamma \simeq 3 \times 10^4$; 
on requiring $\gamma \sigma_0^2 = M_{\rm pl}^2$, it is clear that in general 
IG can incorporate the GUT symmetry breaking scale for inflation. 

In Figure (\ref{rtest}) we compared the WMAP5+BAO+SN Ia
constraints at the 68\% (dotted contour) and the 95\% 
(continuous contour) confidence levels \cite{komatsu} with the prediction 
for the potentials (\ref{CW},\ref{LG}) both in IG (continuous lines) and 
in EG (dashed lines) framework: each point represents a different choice 
for $\gamma$ (or $\xi\equiv\sigma_{0}^{2}/\M^{2}$ in EG) and 
$N_{*}$ (identical markers, from left to right, identify $N_{*}=50,60,70$). 
We observe that in IG for both the above potentials, the small field 
inflation with $\gamma\ge 10^{-3}$ is disfavoured by observations confirming 
what was found in the $n_{s}$ analysis. Again large field inflation in IG fits observations independently of $\gamma$ in contrast with the same regime in EG which requires $\xi\ge10^{3} $. It is however worth noting that both 
the potentials (\ref{CW},\ref{LG}), although very similar to a simple quartic potential in this regime, fit observations well, independently of 
$\mu$ and $\sigma_{0}$, in contrast with such a potential in EG which leads to results lying far away from the 95\% region in the same e-folds interval. 

Let us now study the oscillating regime occuring for $\sigma \simeq \sigma_0$ and consider the symmetry-breaking potential in Eq.
(\ref{LG}) for the sake of simplicity.
We find an approximate analytic solution given by:
\ba
\sigma(t)&=&\sigma_0 + \frac{2}{t} \sqrt{\frac{\gamma}{3\mu}}
\sin \left( \omega t \right) +
\mathcal{O}\left(\frac{1}{t^2}\right) \\
H (t) &\simeq& \frac{2}{3t}\left[1-\sqrt{\frac{6\gamma}{1+6\gamma}}
\cos \left( \omega t \right) \right] + \mathcal{O}\left(\frac{1}{t^2}\right) \,,
\label{coherent}
\ea
where $\omega = \sigma_0 \sqrt{2 \mu/(1+6\gamma)}$.
On fixing the Planck mass after inflation by
$M_{\rm pl}^2 = \gamma \sigma_0^2$ and requiring curvature perturbations
in agreement with observations, the frequency of oscillations $\omega$ is
$\simeq 2.5 \times 10^{13}$ GeV
for $\gamma \ll 1$ (and twice as large for $\gamma \gg 1$).
For the regime of small oscillations this study is equivalent to a mass term obtained from the oscillations of $\sigma$ around a minimum of a potential with $V(\sigma_0) =0$.

Eqs. (\ref{coherent}) can be used to study the evolution of $\delta \sigma$
during the coherent oscillations of $\sigma$: on using Eq. (\ref{scperteq}) the variable
$\tilde{\delta \sigma_k} = \sqrt{a^3 Z} \delta \sigma_k$ can be recast in a
Mathieu-like form:
\be
\frac{d^2 \tilde{\delta \sigma_k}}{d (\omega t)^2} + \left[ A(t) + 2 q_1(t) \sin \left( 2 \omega t \right)
+ 2 q_2 (t) \sin \left( \omega t \right) \right] \tilde{\delta \sigma_k} = 0
\label{mathieu}
\ee
analogously to the EG case \cite{FB}. The leading terms in $q_1$ and $q_2$ originate from the metric and the potential
contributions respectively, and both decay as $1/t$, further $A(t) = k^2/(a^2 \omega^2) + 1 + {\cal O}(t^{-2})$. Because of the two
oscillating terms the time dependent frequency in Eq. (\ref{mathieu}) leads to beats. 
We obtain $q_2 = \sqrt{27 \gamma/(2 (1+6\gamma))}/(\omega t)$ and $q_1 = 2 / (\omega t)$ 
(the latter is the same as the EG case \cite{FB}).

The consequence of coherent oscillations on $\tilde{\delta \sigma_k}$
differs from the resonance for test scalar fields coupled to a massive inflaton \cite{KLS}, 
for which $q(t)$ ($\sim {\cal O}(t^{-2})$) decays 
faster than $k^2/a^2$ and as a consequence the resonance shuts-off asymptotically for large times although it started in the broad regime. 
According to Eq. (\ref{mathieu}), all wavelengths end asymptotically in the first resonance band since $q(t)$ decays more slowly
than $k^2/a^2$. Thus for short-scale modes with $H \ll k/a \ll \omega$, 
$\delta \sigma_k$ oscillates with a constant amplitude instead of decaying \cite{CFTV}. 
Note that such a gravity mediated self-decay 
of the inflaton in its short-scale fluctuations also exists for $V(\phi) = m^2 \phi^2/2$ or 
for a potential with a symmetry-breaking term in EG \cite{explanation_reh}. Although oscillating 
terms decaying as $1/t$ appear in the equation for gravitational waves - on rewriting Eq. (9) in 
terms of the variable $a^{3/2} \sigma h_k$ - such terms do not affect the standard behaviour of short-scale gravitational waves.\\
It is also interesting to study a possible connection of this
inflationary stage with the problem of dark energy, which can also be
modelled in IG through simple potentials \cite{CV,FTV}. Work in this direction
is in progress \cite{CFTV}.

{\bf Acknowledgement.} FF wishes to thank Jerome Martin and 
Karsten Jedamzik for discussions on preheating.


\begin{thebibliography}{99}
\bibitem{explanation_reh}
In a Mathieu-like equation such as Eq. (\ref{mathieu}), a growing solution may be written as 
$\delta \tilde \sigma_k \propto e^{\int^t d (\tau) \mu (\tau)}$ where $\mu (\tau)$ is the ``generalized'' Floquet exponent (time dependent) and can be analyzed through a multiple time scale method \cite{CFTV}.
For the massive case studied in \cite{FB} where $q_2=0 \,, \omega = m$, on evaluating the generalized Floquet exponent in the  
first resonant band for $q_{1} \ll 1$ as $\mu = q_{1}/2 = (mt)^{-1}$ (valid for large scales or 
asymptotically for short scales, i.e. when $A \simeq 1$), we obtain $\tilde Q \sim t$ as was found numerically in 
\cite{FB}. The same conclusion is reached from the conservation of curvature perturbations on large scales.
The same solution holds here for $q_1 \gg q_2$ and this explains the oscillations (with constant amplitude) which occur
for gauge-invariant short-scale inflaton fluctuations $\delta \sigma_k$ at late times.
\end{thebibliography}
\end{document}